\documentclass[prb,twocolumn,showpacs,floatfix,superscriptaddress,epsf,psfig]{revtex4}
\usepackage{epsf}
\usepackage{color}
\usepackage{graphicx}
\usepackage{amsmath, amssymb}
\usepackage{lscape}
\usepackage{rotating}
\begin{document}
\title{Direct Band Gaps in Group IV-VI Monolayer Materials: \\Binary Counterparts of Phosphorene} 
\author{C. Kamal}
\affiliation{Indus Synchrotrons Utilization Division, Raja Ramanna Centre for Advanced Technology, Indore 452013, India}
\author{Aparna Chakrabarti}
\affiliation{Indus Synchrotrons Utilization Division, Raja Ramanna Centre for Advanced Technology, Indore 452013, India}
\affiliation{Homi Bhabha National Institute, Raja Ramanna Centre for Advanced Technology, Indore 452013, India}
\author{Motohiko Ezawa}
\affiliation{Department of Applied Physics, University of Tokyo, Hongo 7-3-1, 113-8656, Japan}

\begin{abstract}
We perform systematic investigation on the geometric, energetic  and electronic properties of group IV-VI binary monolayers (XY), which are the counterparts  of phosphorene, by employing  density functional theory based electronic structure calculations.  For this purpose, we choose the binary systems XY consisting of equal numbers of  group IV (X = C, Si, Ge, Sn) and group VI elements (Y = O, S, Se, Te)  in three geometrical configurations, the puckered, buckled and planar structures.  The results of binding energy calculations show that all the binary systems studied are energetically stable. It is observed that,  the puckered structure, similar to that of phosphorene, is the energetically most stable geometric configuration. Moreover, the binding energies of buckled configuration are very close to those of the puckered configuration.  Our results of electronic band structure predict that  puckered SiO and CSe  are direct band semiconductors with gaps of 1.449 and 0.905 eV, respectively.  Band structure of CSe closely resembles that of phosphorene. Remaining group IV-VI binary monolayers in the puckered configuration and all the buckled monolayers are also semiconductors, but with  indirect band gaps.  Importantly, we find that the difference between indirect and direct band gaps  is very small for many puckered monolayers.  Thus, there is a possibility of making these systems undergo  transition from indirect to direct band gap semiconducting state by a suitable external influence. Indeed, we show in the present work that seven binary monolayers namely SnS, SiSe, GeSe, SnSe, SiTe, GeTe and SnTe become direct band gap semiconductors when they are subjected to a small mechanical strain ($\leq$ 3 $\%$).  This makes nine out of sixteen binary monolayers studied in the present work  direct band gap semiconductors. Thus, there is a possibility of utilizing these binary counterparts of phosphorene in future light-emitting diodes and solar cells.   
\end{abstract}

\pacs{  68.65.-k, 61.46.-w,  81.07.-b, 31.15.E-, 71.15.Mb}
\maketitle


\section{Introduction}
A search of novel monolayer materials is one of the important goals of  material science and condensed matter physics research activity.  Graphene is considered as the most fundamental  two-dimensional (2D) monolayered material and it is one of the well studied 2D systems, both in theory and experiments, due to its many interesting novel properties and potential applications\cite{NetoRev,KatsText}. However, the major disadvantage with graphene is that it has no electronic band gap and hence it is difficult to use graphene in semiconductor device applications.  Recently, the other group IV 2D materials such as silicene, germanene and stanene have attracted great interest \cite{GLay,Kawai,Takamura,LiuPRL,Falko,EzawaNJP,Kamal, hybrid, multilayer},
but it is also difficult to use them as semiconductor devices.
Phosphorene, a monolayer  of phosphorus, has opened up  the field of group V based 2D monolayer materials%
\cite{Li,Reich,Fei,Peng,Rodin,Qiao}. It has an appropriate band gap for electronics applications and is shown to act as a field effect transistor\cite{Li}.  At present, the field is rapidly expanding ever since the experimental realizations of phosphorene\cite{Li,Liu,Xia,Gomez,Koenig,Bus,LiNN}.  In the search of the semiconducting monolayer,  the focus is shifted to group V based systems, namely, nitrogenene, arsenene, and antimonene, which are nitrogen, arsenic and antimony based monolayers respectively. They have also been predicted to be  stable by first-principles calculations\cite{As,Ange,ZhuAs,WangAs,AsR, Ciraci-nitrogene}.  In case of group III monolayers,  planar aluminene,  monolayer of aluminum, is predicted to be stable but it is a metal\cite{aluminene}. 

Recently,  a binary 2D system consisting of an equal number of two different elements draws growing attention. The properties of the binary 2D systems can be entirely different from those of elemental 2D systems. For instance, hexagonal boron nitride (h-BN), which has a planar honeycomb structure similar to graphene, shows insulating behaviour with a large band gap in contrast to the semi-metallic behaviour of graphene. In general, there are several possible materials, BN, AlN, GaN, InN, BP, AlP, GaP, InP, BAs, AlAs, GaAs, etc,  consisting of equal numbers of group III and group V atoms.  These binary systems have the same number of valence electrons in an unit cell as that of graphene, silicene, germanene and stanene. Hence, it can be considered as derivatives of group IV monolayers. 
It has been observed in previous studies that all the group III-V  binary counterparts of graphene, studied up to now,  show semiconducting behaviour\cite{Sahin}.

Now, it is natural to ask  questions: whether there exist hexagonal monolayers made up of  group IV and VI elements, which are counterparts of phosphorene, arsenene, antimonene or in general, monolayer derivatives  of  group V. If they exist, what are their properties?  From both fundamental and application points of view, it is important to perform studies on these group IV-VI binary systems.
Although there are some first-principles calculations on SnS, SnSe, GeS, and GeSe\cite{Gomes,GuoPRB,Ding,RodinC}, 
all of them are known to be indirect-gap semiconductors. 
Direct gap semiconducotors are desirable for practical applications. 
In this paper, we systematically investigate the geometric, energetic and electronic properties of  group IV-VI monolayers (XY) (with X = C, Si, Ge, Sn  and  Y = O, S, Se, Te)  by employing density functional theory (DFT) based electronic structure calculations.  

This choice of elements would give sixteen possible combinations of materials, namely, CO, SiO, GeO, SnO, CS, SiS, GeS, SnS, CSe, SiSe, GeSe, SnSe, CTe, SiTe, GeTe, and SnTe.  For each of these 2D binary monolayers, we have considered three different possible geometrical configurations such as the puckered, buckled and planar structures.  The binding energy calculations predict that all the sixteen binary monolayers studied in the present work  are energetically stable. For most of the cases, the puckered configuration, similar to that of phosphorene, is the most stable configuration.  Among the minimum energy configurations, we observe that two systems  SiO and CSe in the puckered configuration are direct band semiconductors with  band gaps of 1.449 and 0.905 eV, respectively, while the remaining materials are indirect band gap semiconductors.  Moreover, our calculations predict that seven monolayers (SnS, SiSe, GeSe, SnSe, SiTe, GeTe and SnTe) in the puckered configuration undergo  an indirect-to-direct band gap transition by the application of small mechanical strain ($\leq$ 3 $\%$).  These results indicate that  the group IV-VI 2D materials are promising materials for  applications in light-emitting devices and solar cells.

This paper is organized in the following manner. In the next section, we describe the computational details employed in the
present work. Section III contains the results and discussion, and then in Section IV, we give summary of our results.

\section{Computational Details}

\quad
Density functional theory (DFT)\cite{dft} based calculations have been performed using Vienna ab-initio simulation package (VASP)\cite{vasp} within the framework of the projector augmented wave (PAW) method. We employ generalized gradient approximation (GGA) given by Perdew-Burke-Ernzerhof (PBE)\cite{pbe} for exchange-correlation functional.  The plane waves are expanded with energy cut of 400 eV. We use Monkhorst-Pack scheme for k-point sampling of  Brillouin zone integrations with 41$\times$31$\times$1  and 31$\times$31$\times$1 for the puckered and buckled/planar configurations, respectively.   The convergence criteria for energy in SCF cycles is chosen to be 10$^{-6}$ eV.   The geometric structures are optimized by minimizing the forces on individual atoms with the criterion that the total force on each atom is below 10$^{-2}$ eV/\AA{}. We choose the following valence electronic configurations for C [2s$^2$ 2p$^2$], O [2s$^2$ 2p$^4$], Si [3s$^2$ 3p$^2$], S [3s$^2$ 3p$^4$], Ge [4s$^2$ 3d$^{10}$ 4p$^2$], Se [4s$^2$ 4p$^4$], Sn [5s$^2$ 4d$^{10}$ 5p$^2$] and Te [5s$^2$ 5p$^4$]. In order to mimic the two-dimensional system, we employ a super cell geometry with a vacuum of about 18 \AA {} in the direction perpendicular to the plane of 2D sheet so that the interaction between two adjacent unit cells in the periodic arrangement is negligible. The geometric structures are drawn using XCrySDen software\cite{xcrysden}

\section{Results and Discussions}
 \subsection{Geometric Structure and Binding Energy}

\begin{figure}[!t]
\begin{center}
\includegraphics[width=0.45\textwidth]{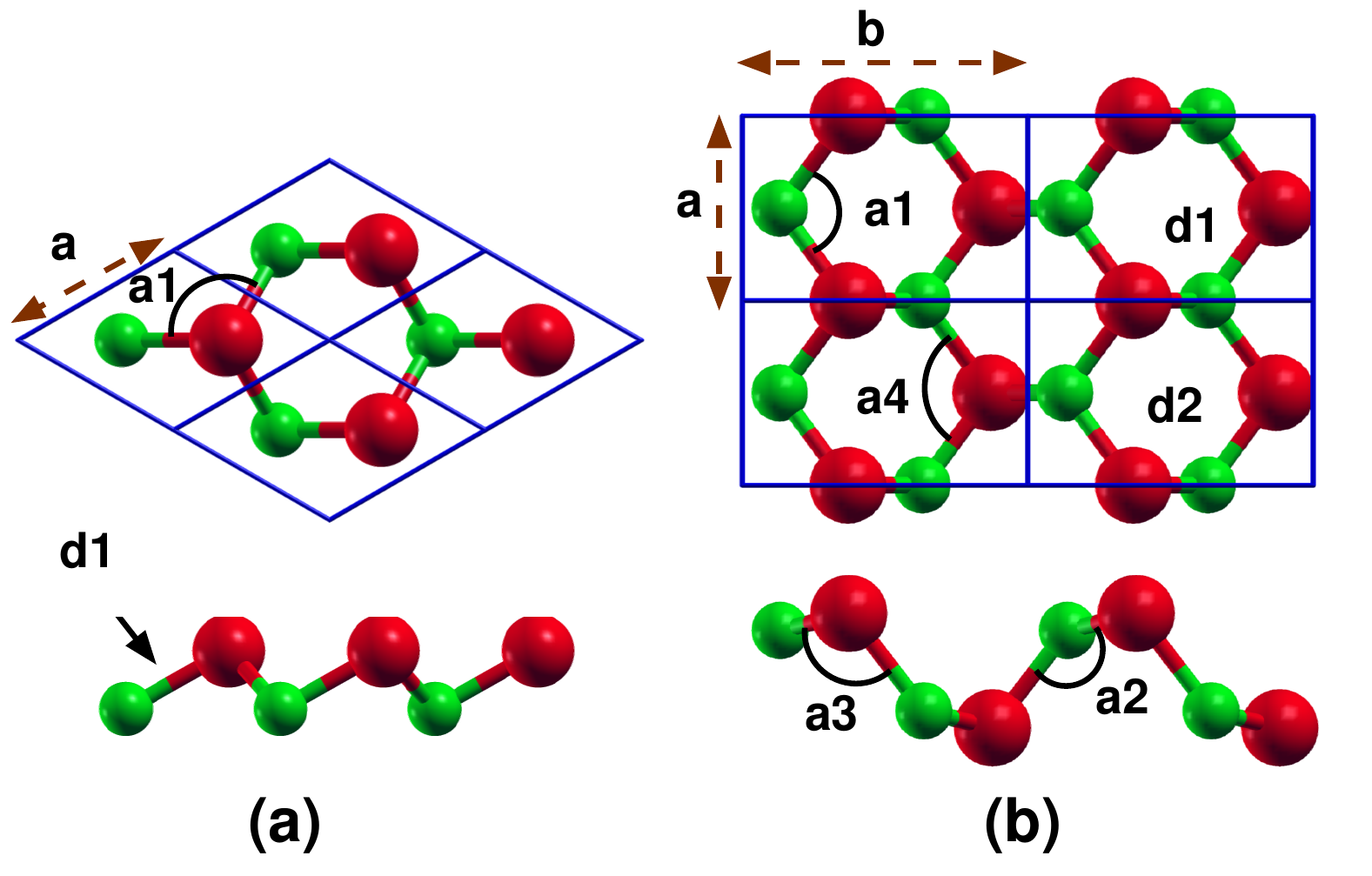}
\end{center}
\caption{ (color online) The geometric structures of group IV-VI binary monolayer in the (a) buckled and (b) puckered configurations. Small green (big red) balls represent group IV (group VI) atoms.}
\label{FigStruct}
\end{figure}

We consider the hexagonal monolayer structure made of group IV and VI elements in three different possible geometrical configurations, the planar, buckled and puckered structures. These three geometrical configurations are found to be the minimum energy structures of  already known monolayer systems. Graphene and hexagonal boron-nitride  form the planar structure, while silicene, germanene and stanene possess the buckled structure. Recently fabricated phosphorene has been shown to form the puckered structure. In addition,  buckled phosphorene, which is  named as blue phosphorene,  has also been proposed theoretically\cite{Blue}.  In the case of arsenene, it is observed that both of the puckered and buckled structures are stable and their binding  energy difference is small\cite{As}.  

The hexagonal structure is bipartite consisting of the A and B sublattices.  We assume that the group IV (X = C, Si, Ge, Sn)  atoms reside at the A sites, while the group VI (Y = O, S, Se, Te) atoms reside at the B sites forming the hexagonal structure. The ball and stick model of group IV-VI monolayers in the buckled and puckered configurations are shown in Fig.1.  The unit cell of the buckled structure contains one group IV atom and one group VI atom, while that of the puckered structure contains two group IV atoms and two group VI atoms as shown in Fig.\ref{FigStruct}.  

The results of the binding energy and geometric analysis of group IV-VI binary monolayers in these three configurations are summarized in Table I. It is observed that the puckered structure is slightly deformed as compared with phosphorene since the two sublattices are not the same.  In order to understand how the geometrical properties vary with the materials, we have also plotted variations of (a) the binding energy, (b) the lattice constant, (c) the bond length and (d) the bond angle of the binary monolayer XY in Fig.2.

\begin{table}[]
\footnotesize
\begin{center}
\caption{The results of binding energy, geometrical parameters and band gaps for the group IV-VI binary monolayers in three different geometrical configurations, (a) puckered (Pmn2$_1$), (b) buckled (P3m1) and (c) planar (P6/mmm). }
 \begin{tabular}{lcccccccc}
\hline
\hline

	&		Binding 	&	\multicolumn{2}{c}{Lattice }&		Bond &\multicolumn{3}{c}{	Band Gap }			\\  \cline{6-8}
XY	      &	Energy	&	\multicolumn{2}{c}{Constant (\AA{})}	&	Length	  & Indirect	&	Direct	&	Diff	\\ \cline{3-4}
	      &		(eV/atom)	&	a&	b	&	(\AA{})	  &  &	(eV)	&	\\

\hline
(a)\\
CO	&	-5.07	&	2.180	&	4.010	&	1.379	&	1.370	&	1.631	&	0.262	\\
SiO	&	-5.27	&	2.739	&	4.701	&	1.843, 1.859	&	-	&	1.449 	&	-	\\
GeO	&	-4.67	&	3.055	&	4.801	&	1.956, 1.986	&	2.759	&	2.963	&	0.204	\\
SnO	&	-4.45	&	3.400	&	4.764	&	2.127, 2.163	&	2.682	&	2.787	&	0.105	\\
CS	&	-4.37	&	2.795	&	4.323	&	1.757, 1.849	&	0.962	&	1.192	&	0.231	\\
SiS	&	-3.91	&	3.352	&	4.774	&	2.300, 2.344	&	1.423	&	1.550	&	0.127	\\
GeS	&	-3.66	&	3.642	&	4.492	&	2.462, 2.423	&	1.757	&	1.856	&	0.099	\\
SnS	&	-3.55	&	4.047	&	4.347	&	2.728, 2.595	&	1.447	&	1.644	&	0.197	\\
CSe	&	-3.84	&	3.034	&	4.299	&	1.961, 2.014	&	-	&	0.905 	&	- 	\\
SiSe	&	-3.53	&	3.737	&	4.400	&	2.524, 2.448	&	0.673	&	0.959	&	0.287	\\
GeSe	&	-3.37	&	3.965	&	4.302	&	2.661, 2.544	&	1.145	&	1.160	&	0.015	\\
SnSe	&	-3.28	&	4.260	&	4.453	&	2.887, 2.730	&	0.929	&	1.025	&	0.096	\\
CTe	&	-3.68	&	3.390	&	3.889	&	2.164, 2.181	&	0.546	&	1.247	&	0.702	\\
SiTe	&	-3.23	&	4.109	&	4.300	&	2.641, 2.772	&	0.395	&	0.466	&	0.072	\\
GeTe	&	-3.09	&	4.238	&	4.376	&	2.736, 2.883	&	0.850	&	0.906	&	0.056	\\
SnTe	&	-2.99	&	4.542	&	4.581	&	2.931, 3.164	&	0.666	&	0.699	&	0.033	\\
\hline
(b)\\
CO	&	-4.34	&	2.454	&-		&	1.636	&	3.284	&	3.493	&	0.281	\\
SiO	&	-5.02	&	2.815	&-		&	1.884	&	0.706	&	1.069	&	0.363	\\
GeO	&	-4.37	&	3.124	&-		&	2.032	&	2.232	&	2.582	&	0.350	\\
SnO	&	-4.17	&	3.442	&-		&	2.204	&	1.638	&	2.122	&	0.484	\\
CS	&	-4.08	&	2.836	&-		&	1.880	&	1.366	&	1.698	&	0.332	\\
SiS	&	-3.92	&	3.299	&-		&	2.321	&	2.191	&	2.493	&	0.302	\\
GeS	&	-3.64	&	3.485	&-		&	2.428	&	2.490	&	2.712	&	0.222	\\
SnS	&	-3.47	&	3.757	&-		&	2.616	&	2.315	&	2.602	&	0.286	\\
CSe	&	-3.64	&	3.063	&-		&	2.055	&	1.547	&	1.762	&	0.216	\\
SiSe	&	-3.55	&	3.521	&-		&	2.477	&	2.124	&	2.327	&	0.203	\\
GeSe	&	-3.36	&	3.676	&-		&	2.568	&	2.278	&	2.479	&	0.202	\\
SnSe	&	-3.22	&	3.916	&-		&	2.747	&	2.219	&	2.446	&	0.227	\\
CTe	&	-3.43	&	3.348	&-		&	2.231	&	1.283	&	1.564	&	0.281	\\
SiTe	&	-3.20	&	3.835	&-		&	2.690	&	1.833	&	2.011	&	0.177	\\
GeTe	&	-3.06	&	3.939	&-		&	2.768	&	1.728	&	1.978	&	0.251	\\
SnTe	&	-2.93	&	4.151	&-		&	2.947	&	1.790	&	2.135	&	0.346	\\
\hline
(c)\\
CO	&	-3.50	&	2.923	&-		&	1.688	&	-	&	-	&	-	\\
SiO	&	-4.26	&	3.461	&-		&	1.998	&	0.035	&	0.098	&	0.063	\\
GeO	&	-4.00	&	3.653	&-		&	2.109	&	0.675	&	0.685	&	0.010	\\
SnO	&	-3.89	&	3.924	&-		&	2.265	&	0.818	&	0.879	&	0.061	\\
CS	&	-3.58	&	3.157	&-		&	1.823	&	-	&	-	&	-	\\
SiS	&	-3.24	&	4.157	&-		&	2.400	&	-	&	-	&	-	\\
GeS	&	-3.13	&	4.377	&-		&	2.527	&	-	&	-	&	-	\\
SnS	&	-3.07	&	4.694	&-		&	2.710	&	- & 0.142	&	-	\\
CSe	&	-3.27	&	3.546	&-		&	2.047	&	-	&	-	&	-	\\
SiSe	&	-3.00	&	4.383	&-		&	2.531	&	-	&	-	&	-	\\
GeSe	&	-2.91	&	4.559	&-		&	2.632	&	-	&	-	&	-	\\
SnSe	&	-2.84	&	4.888	&-		&	2.822	&	-	&	-	&	-	\\
CTe	&	-3.13	&	3.846	&-		&	2.221	&	-	&	-	&	-	\\
SiTe	&	-2.70	&	4.711	&-		&	2.720	&	-	&	-	&	-	\\
GeTe	&	-2.62	&	4.865	&-		&	2.809	&	-	&	-	&	-	\\
SnTe	&	-2.54	&	5.213	&-		&	3.010	&	-	&	-	&	-	\\
\hline
\hline
\end{tabular}
\end{center}
\end{table}

\begin{figure*}
\begin{center}
\includegraphics[width=1.0\textwidth]{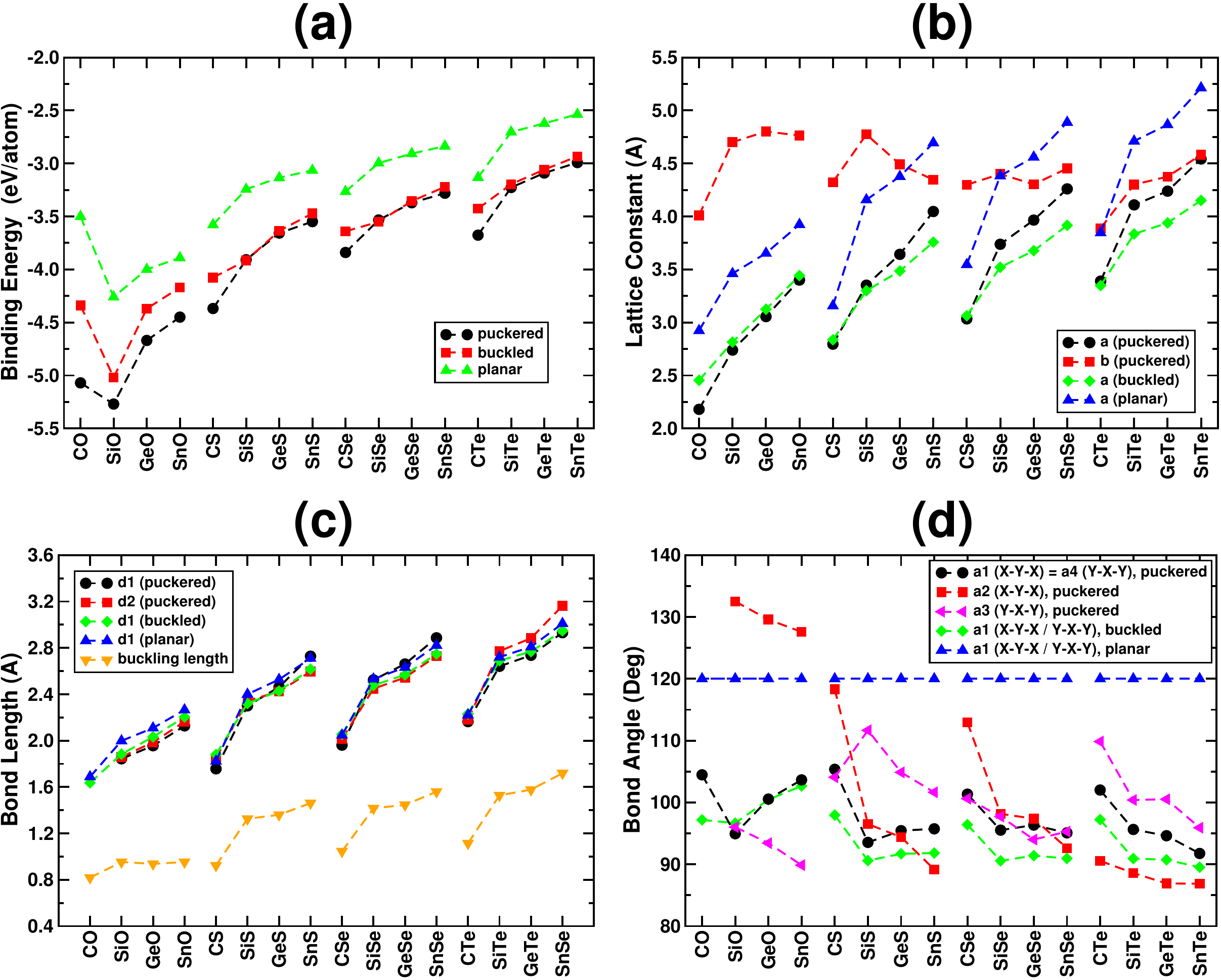}
\end{center}
\caption{(color online) Variations of the binding energy and the geometrical properties of the hexagonal group IV-VI binary monolayers.
(a) The binding energies  for the three geometric configurations, the planar, puckered and buckled structures.  The puckered structure is the minimum energy configuration except for SiS and SiSe systems. However, the binding energy difference between the puckered and buckled structure is very small. 
(b) The lattice constants for the IV-VI monolayers. There are two lattice constants for the puckered structure since the structure is anisotropic.
(c) The bond length for the IV-VI monolayers. There are two different types of X-Y bond in the puckered structure. Bond lengths $d1$ and $d2$ are illustrated in Fig.\protect\ref{FigStruct}.  It also contain the buckling length  for buckled structure.
(d) The bond angle for the IV-VI monolayers.  There are four types of bond angles, $a1,a2,a3$ and $a4$ in the puckered structure
(illustrated in Fig.\protect\ref{FigStruct}.)
}
\label{FigProperties}
\end{figure*}

\begin{figure}[!t]
\begin{center}
\includegraphics[width=0.45\textwidth]{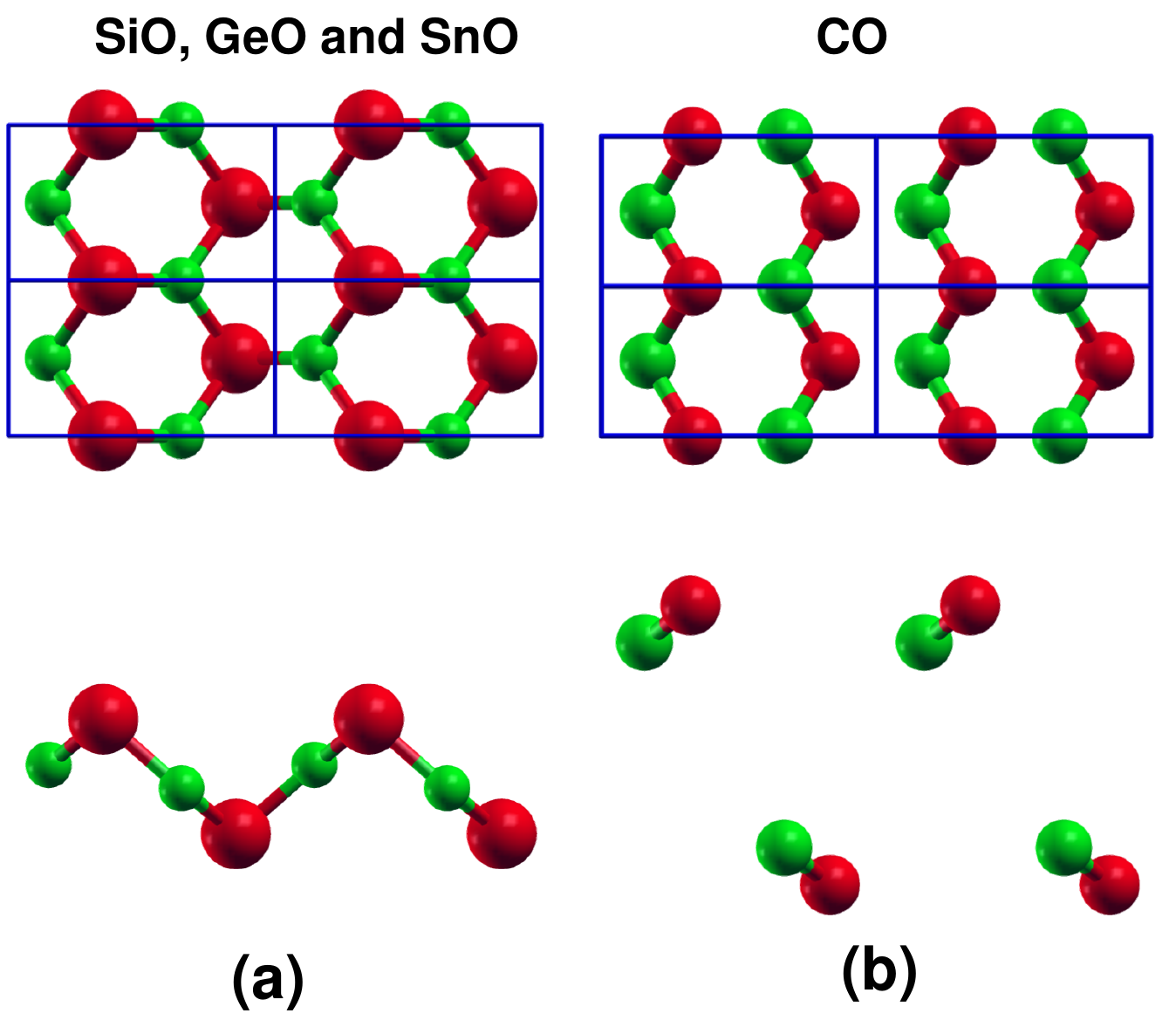}
\end{center}
\caption{(color online) The optimized geometric structures of oxides: (a) SiO, GeO, SnO (b) CO in the puckered configuration. CO does not form a 2D structure but a liner chain along the lattice $a$.  However, the other oxides form 2D structure which is slightly different from the puckered configuration. Small green (big red) balls represent group IV (group VI) atoms).}
\label{FigStruct10}
\end{figure}

The binding energy ($E_B$) of the group IV-VI binary monolayers (XY) is calculated by using the formula
\begin{equation}
 E_B = E_{XY(2D)} - n \left[ E_{X(\text{atom})} + E_{Y(\text{atom})}\right],
\end{equation}
where $E_{XY(2D)}$, $E_{X(\text{atom})}$, $E_{Y(\text{atom})}$ are the energies of group IV-VI binary monolayers, their constituent atoms X and Y, respectively; $n$ is the number of group IV  or VI atoms in the unit cell.  Then, we scaled the binding energy by the number of atoms in the unit cell.  It is observed from Table I that all the binary monolayers studied in the present work are energetically stable (the negative sign of  $E_B$ indicates that they form bound states).  

Our calculations predict that the puckered structure is the minimum energy configuration for group IV-VI binary monolayers, except for SiS and SiSe.  For many monolayers, we find that the binding energies of the buckled structure are very close to those of the puckered structure.  It is important to note that even for SiS and SiSe, the binding energy differences between the buckled and puckered configurations are 0.01 and 0.02 eV/atom, respectively, which are very small. Thus, both the geometric configurations of SiS and SiSe are nearly equally probable at room temperature. 

On the other hand, the planar structure is the least stable structure for group IV-VI binary monolayers.  The hexagonal planar structure  can support  the sp$^2$ hybridization, whereas the favourable hybridization in group V monolayers (phosphorene and arsenene) is sp$^3$.  Thus, our results suggest that the hybridization in group IV-VI binary monolayers is similar to those of phosphorene and arsenene. The hybridization in phosphorene and arsenene is sp$^3$-like but they do not show the characteristic bond angle of 109.47 due to the presence of  non-bonding lone pair of electrons in one of the hybridized orbitals. Ammonia (NH$_3$) is the molecular equivalent for the group IV-VI binary monolayers.  However, we wish to mention here that, unlike the presence of pure covalent-like bonding in elemental group V monolayers,  there must be some amount of ionic component in the bonding in group IV-VI binary monolayers due to the difference in electronegativity between group IV and VI elements. The details of the Bader charge analysis for group IV-VI binary monolayers will be discussed in the next section. 

In Fig.2, we plot the binding energies, lattice constants, bond lengths and bond angles of the hexagonal group IV-VI binary monolayers.  For a given group IV element X, the binding energy increases monotonically when we go from O to Te  in  group VI. The trend is similar for all the group IV elements except for the monolayer CO [See Fig.2 (a)]. 
The reason that CO deviates from the trend is that it forms an one-dimensional structure. There is only one lattice constant ($a$), one bond length ($d1$) and one bond angle ($a1$) in both planar and buckled structures. They have the space group P6/mmm and P3m1, respectively. In case of puckered structure with the space group Pmn2$_1$, there are two lattice constants ($a$, $b$), two bond lengths ($d1$, $d2$) and four bond angles ($a1$, $a2$, $a3$, $a4$), which are shown in Fig.1. We observe that variations of the lattice constant and the bond length are similar to that of the binding energy.  They increase smoothly when we go down the columns of both groups IV and VI. These variations can be explained as follows: As we go down the column either in group IV or VI, the distance of valence electrons from the nucleus increases, and hence the atomic radius of element which is responsible for the binding energy of the system increases with the size.  Both the lattice constants and the bond lengths exhibit similar trend due to the increase in the atomic radii of the elements. We also observe that the buckling length, which is defined as the vertical distance between the atoms at the A and B sites in the buckled structure, increases proportionately with the increase in bond length and hence they show nearly same trend as that of the bond length and lattice constant $a$. 

On the other hand, the bond angles in the buckled and puckered structures of group IV-VI binary monolayer show decreasing trend when we go down the columns of group IV or VI.  Due to the symmetry of the space group, the bond angle in the planar structure is fixed at 120$^\circ$ and the bond angle $a1$ is equal to $a4$ in the puckered structure.  It is important to note that bond angles $a2$ and $a3$ are identical in phosphorene and arsenene, where the atoms are placed in two planes.  The puckered structures of phosphorene and arsenene contain the following bond angles $a1=a4=98.15^\circ$; $a2=a3=103.69^\circ$ and $a1=a4=94.64^\circ$; $a2=a3=100.80^\circ$, respectively. The difference between these two angles ($a2$ and $a3$) indicates that the atoms (X and Y) in the unit cell are not in the same plane. 

It is observed from Fig.2(d) that the puckered structures of group IV-VI monolayers are significantly deformed in comparison with phosphorene. 
A large deformation is observed for the oxides since the difference between $a2$ and $a3$ is very high.  The bond angles in oxides (CO, SiO, GeO and SnO) also show  slightly different trend as compared to the sulfides, selenides and tellurides. The detailed analysis shows that the geometries of the puckered oxides are converged to slightly different structures. The ball and stick models of optimized geometries of oxides are shown in Fig.3.  We observe large values of $a_2$ for SiO, GeO and SnO. Moreover, CO in the puckered form does not form a 2D structure, but forms a one-dimensional chain along the $a$ axis. Also in the case of the buckled structure, the bond angles of the oxides show different trend in comparison with other monolayers. This  clearly indicates that the behaviour of oxides are different from those of sulfides, selenides and tellurides.

\begin{figure*}[]
\begin{center}
\includegraphics[width=0.8\textwidth]{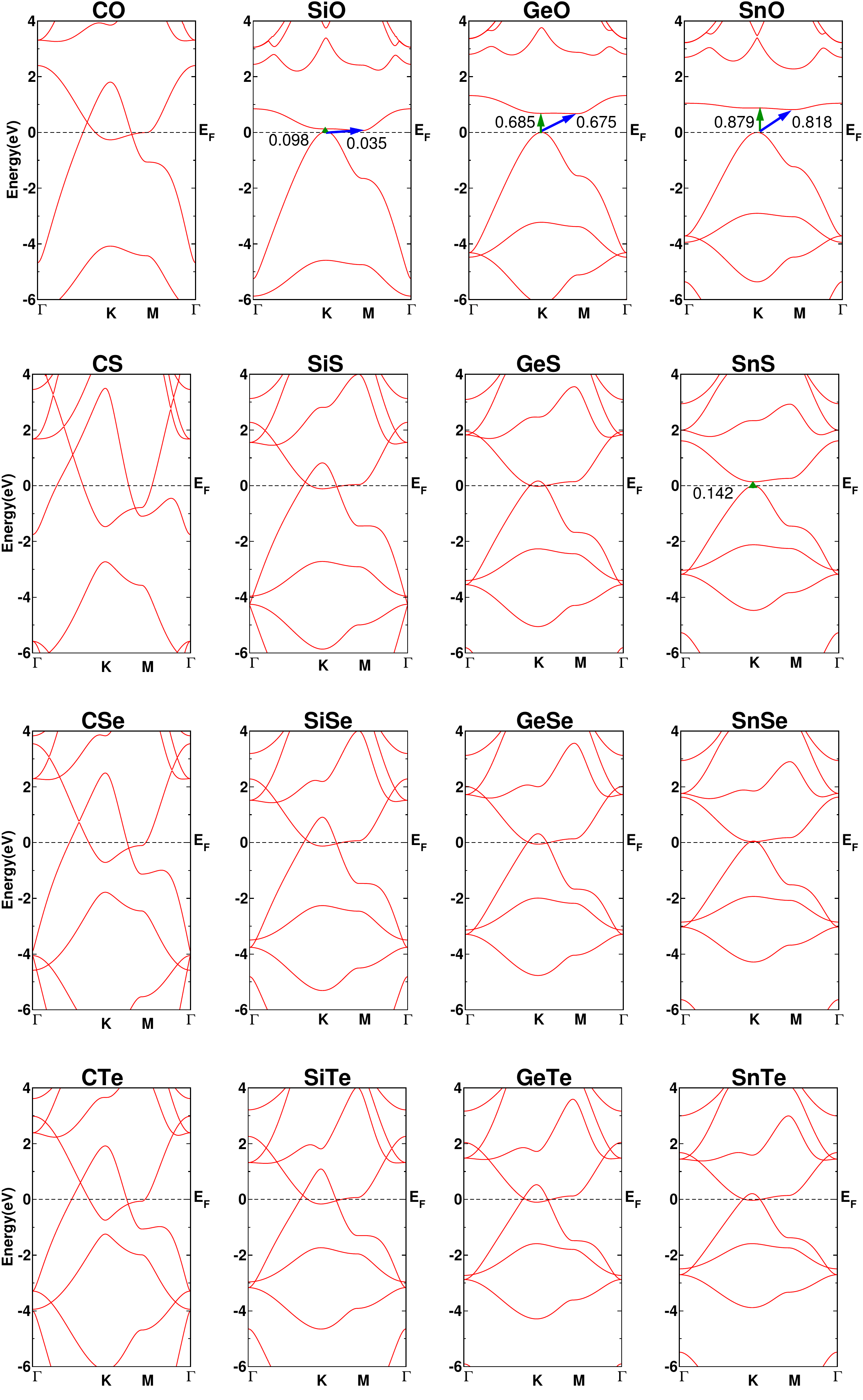}
\end{center}
\caption{(color online) The band structure of group  IV-VI monolayers in the least stable planar configuration.  SnS possesses a direct gap at the $K$ point, while SiO, GeO and SnO are indirect-gap semiconductors. Remaining systems are metal. The indirect band gap is indicated by oblique arrow (green color). The vertical arrows (blue color) indicate the vertical transitions from the local maxima of the valence band to the local minima of the conduction band.  Numerical values written near the arrows represent the corresponding band gaps.}
\label{FigBandStruc3}
\end{figure*}

\begin{figure*}[]
\begin{center}
\includegraphics[width=0.8\textwidth]{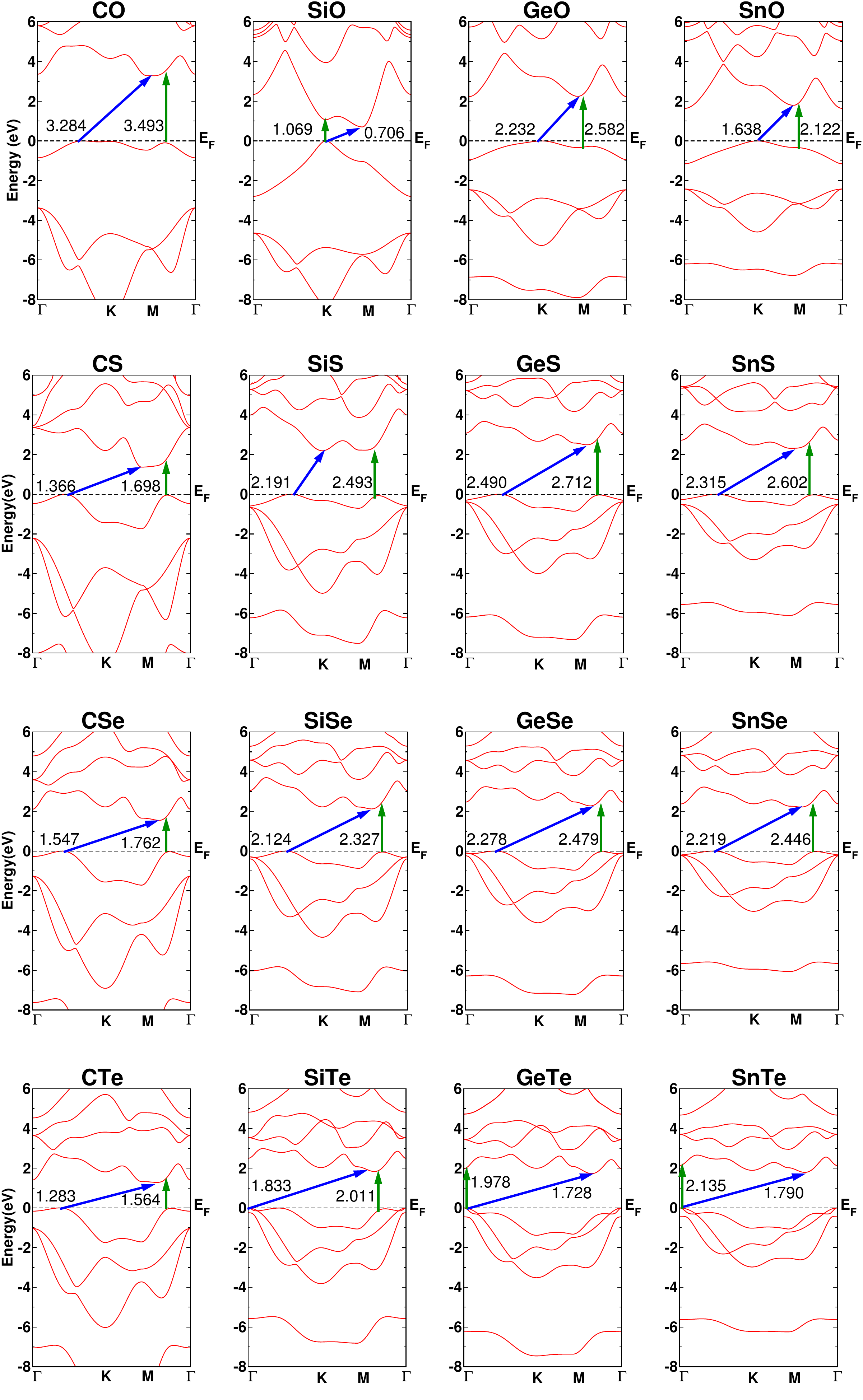}
\end{center}
\caption{(color online) The band structure of group  IV-VI monolayers in the buckled configuration. All of them are indirect-gap semiconductors. The indirect band gap is indicated by oblique arrow (green color). The vertical arrows (blue color) indicate the vertical transitions from the local maxima of the valence band to the local minima of the conduction band. Numerical values written near the arrows represent the corresponding band gaps.}
\label{FigBandStruc2}
\end{figure*}

\begin{figure*}[]
\begin{center}
\includegraphics[width=0.8\textwidth]{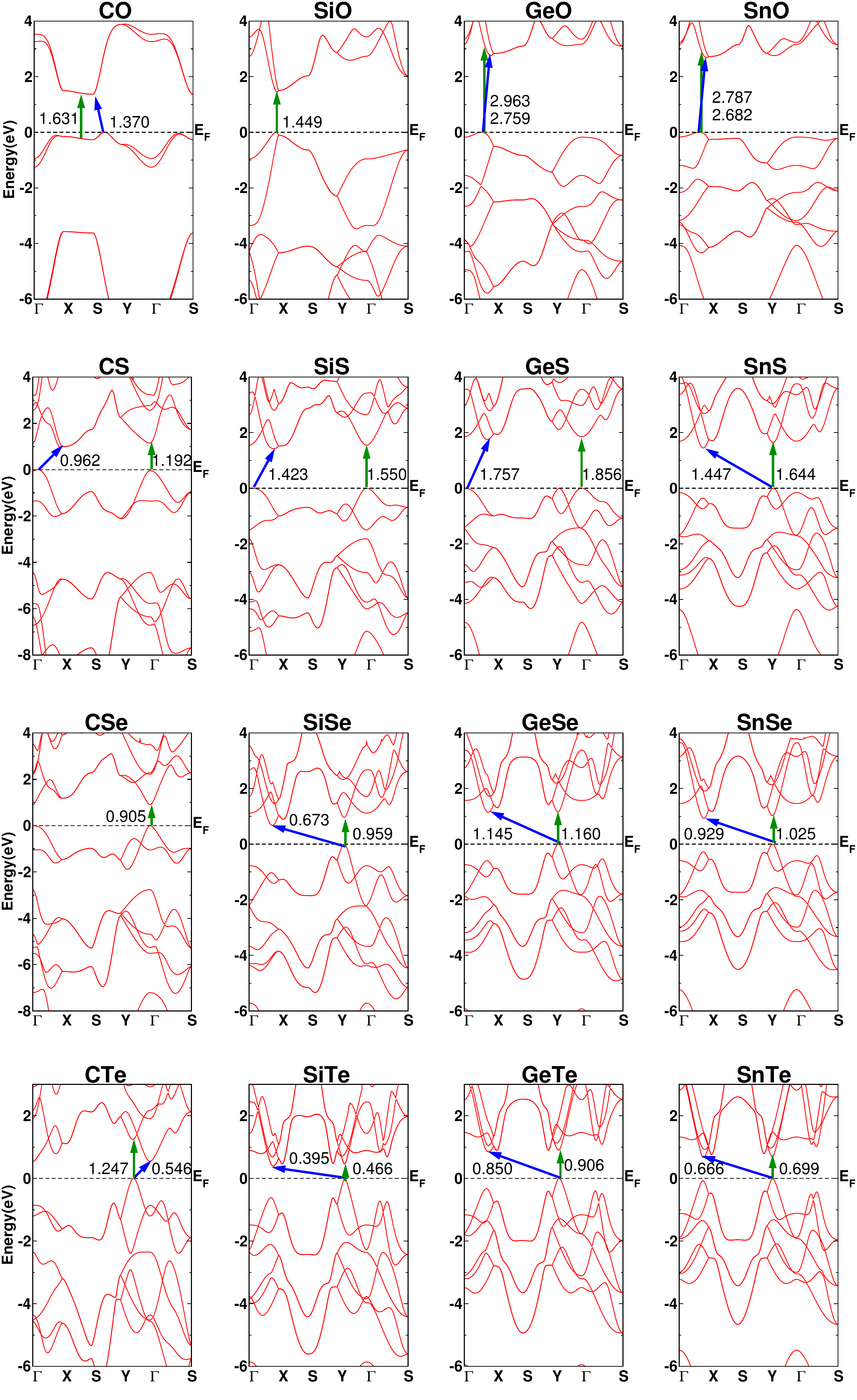}
\end{center}
\caption{(color online) The band structure of group IV-VI monolayers in the puckered configuration. SiO and CSe show  direct gap at the $\Gamma$ point and along the $\Gamma-X$ direction, respectively, while the others are indirect-gap semiconductors. The indirect band gap is indicated by oblique arrow (green color). The vertical arrows (blue color) indicate the vertical transitions from the local maxima of the valence band to the local minima of the conduction band. Numerical values written near the arrows represent the corresponding band gaps.}
\label{FigBandStruc}
\end{figure*}

\subsection{Band Structure and Density of States}
In this subsection, we discuss the results of the electronic band structures and the density of states (DOS) of 2D monolayers made up of  group IV and VI elements. The electronic band structures of  group IV-VI binary monolayers in the planar, buckled and puckered configurations are given in Fig. 4-6, respectively.  It is observed from Fig.4 that most of the group IV-VI monolayers in the least stable planar configuration are metallic due to the strong overlap of the conduction and valence bands. However, the systems made of SiO, GeO, SnO and SnS show the semiconducting behaviour. We represent the indirect and direct band gaps by the oblique (blue color) and vertical (green color) arrows, which are the transitions between the top of the valence band and the  bottom of the conduction band. Since these are the least stable configurations for group IV-VI monolayers, we do not discuss their results in detail.

\begin{figure}[]
\begin{center}
\includegraphics[width=0.4\textwidth]{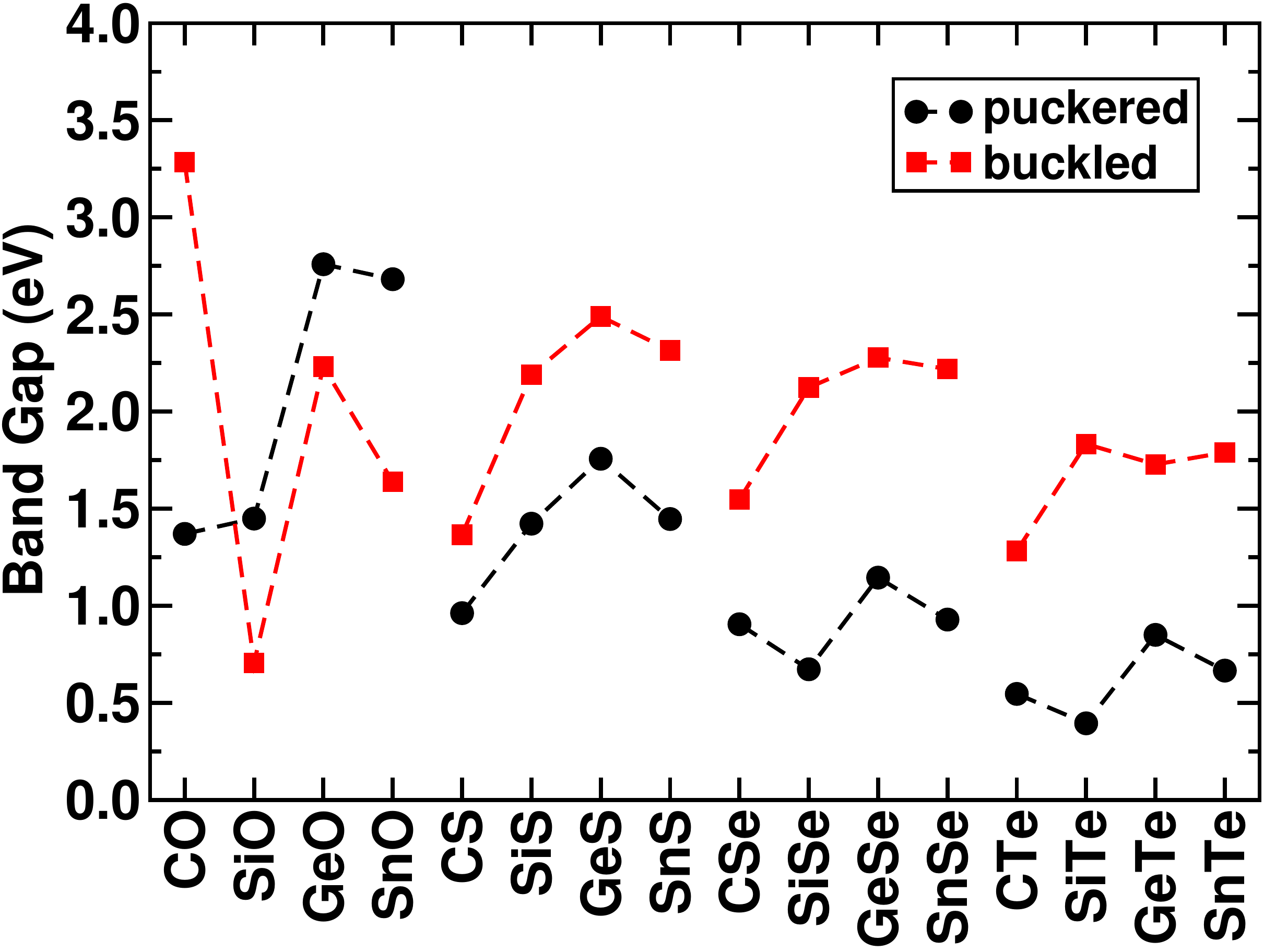}
\end{center}
\caption{(color online) The band gap of the monolayer IV-VI monolayers.}
\label{FigGap}
\end{figure}

Energetically the next stable configuration is the buckled structure.  Similar to the puckered geometric structure, the buckled structure can support the sp$^3$-like hybridization. Thus, the binding energies of the buckled structures are higher than those of the planar structures and they are also closer to those of the puckered structures. The results presented in Fig.5 indicate that all the group IV-VI binary materials are indirect band gap semiconductors.  Moreover, the nature of dispersion in band structures of sulfides, selenides and tellurides looks very similar to those of buckled phosphorene and arsenene\cite{Blue,As}.  Both buckled phosphorene (blue) and arsenene are indirect band gap semiconductors.  Further, it is observed that the band structure of oxides is somewhat different.  In addition, it has been observed that the differences between the fundamental indirect and direct band gaps are very large and they are much higher than the thermal energy (25 meV) at ambient temperature. The lowest difference in the band gaps observed is 177 meV for SiTe. Thus, though the binding energies of buckled structures are closer to those of puckered structures, they may not be useful in any optoelectronic device application due to their indirect band gap semiconducting character. 

Let us now discuss  the results of the band structures of group IV-VI binary monolayers in the most stable puckered structure.  Our calculations predict that all the binary monolayers in the puckered configuration are semiconductors.  It is observed from Fig.4 that  SiO and CSe possess direct band gaps of 1.449 and 0.905 eV, respectively.  It is interesting to note that the band structure of CSe is quite similar to that of phosphorene. In both cases, the direct band gap transitions occur at the $\Gamma$ point and the value of the band gap in CSe is almost the same as that of phosphorene 
($\sim0.91$ eV)\cite{PengP,Liu,TranB}.  Hence,  CSe can be considered as an alternate, group IV-VI based, counterpart of phosphorene.  
The reason that the band structures of phosphorene and CSe are similar is  the electronegativity is same for both C and Se atoms.

On the other hand, the band dispersions observed in SiO and also in other oxides look quite different from those of phosphorene and arsenene.  The observed differences in nature of dispersion in oxides as compared to those of  phosphorene and arsenene can be attributed to the significant deformation in the geometric structures of oxides.  In this case, the direct band gap occurs along  the $\Gamma$-X direction which is different from that of CSe and phosphorene.

In addition,  it is to be noted that the band structure of CO is entirely different as compared to those of other oxides since it does not form complete 2D like geometric structure. Though the other sulfides, selenides and tellurides show indirect band semiconducting behaviour, their band structures are highly anisotropic and the nature of dispersion is nearly similar to those of phosphorene\cite{Liu,PhosA} and arsenene\cite{As}.  

We also plot the fundamental band gaps of the puckered and buckled structures in Fig.\ref{FigGap} which are given in Table I.  Except for oxides,  the band gap of the  puckered structure is smaller than that of the buckled structure for each material.  The trend is consistent with the results observed in phosphorene and arsenene.  Band gap in puckered phosphorene is $0.91$eV\cite{Liu,Peng,Tran} which is smaller than $~ 2$eV of band gap observed in buckled phosphorene\cite{Blue}. In case of arsenene, the values of band gaps in puckered and buckled structures are 0.831 and  1.635 eV respectively\cite{As}. We find that there  is an overall  decreasing trend in the values band gaps of group IV-VI binary monolayers when we go from O to Te.   Before proceeding further, we wish to compare the properties of group IV-VI binary monolayers with those of bulk materials available in the literature in order to understand the variation upon the reduction in dimension. In Table III, we summarize the experimentally obtained geometrical parameters and band gaps for some of the bulk materials, namely GeS, SnS, GeSe and SnSe which are available in the literature. We find that our theoretical results for the lattice constants  ($a$ and $b$, given in Table I) for the puckered monolayers  match well with the corresponding experimental values for these bulk materials. The maximum differences in the values of $a$ and $b$ are found to be less than 4.5 \%.  It is natural to expect very close match between the results of geometric structure of monolayer and bulk material since the latter has layered structure as well.

\begin{figure*}[]
\begin{center}
\includegraphics[width=1.1\textwidth, angle=90]{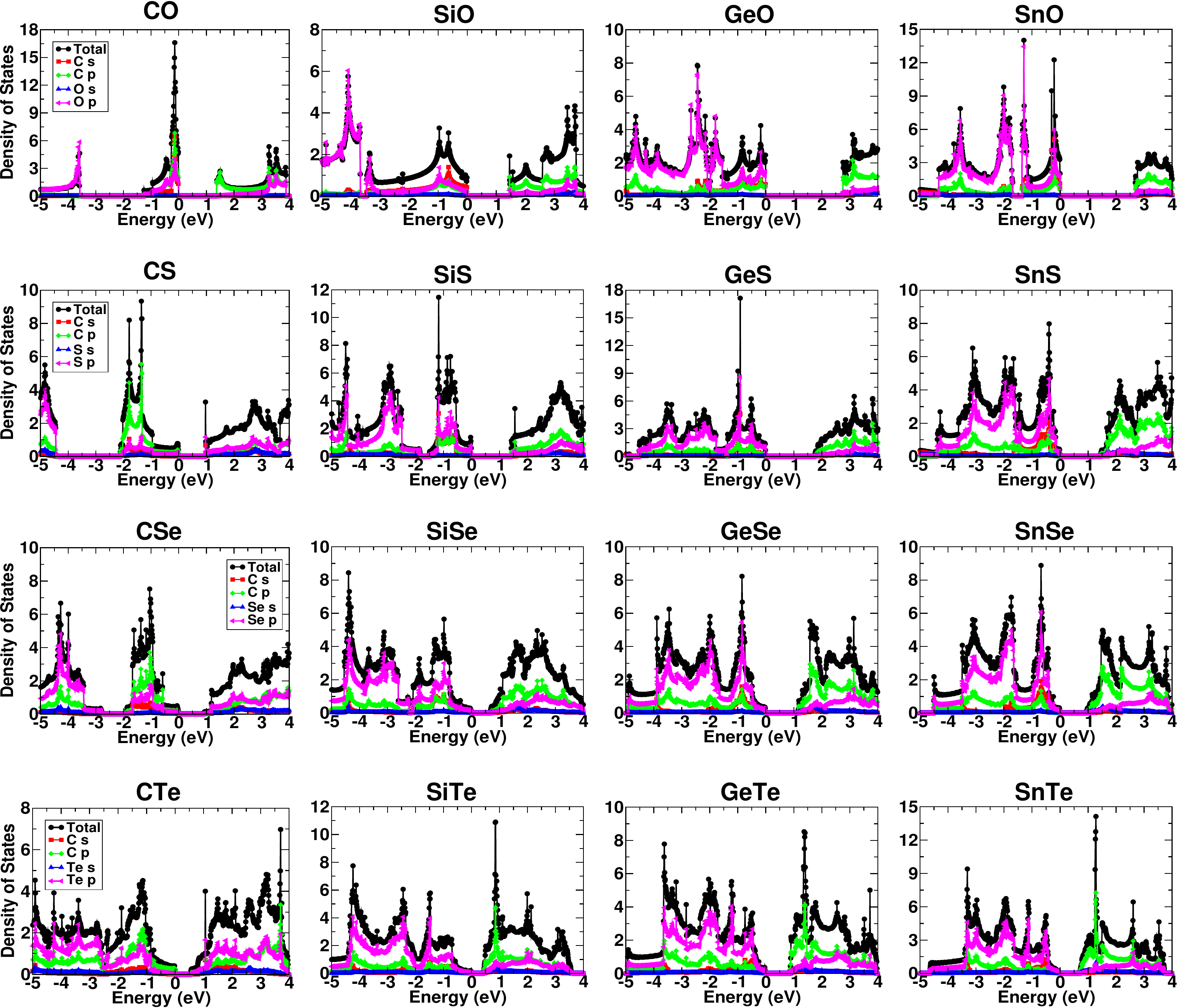}
\end{center}
\caption{ (color online) The total and partial DOS (states/eV/fu) of the puckered IV-VI monolayers.  For better visibility, the values for partial DOS have been scaled by the factor 2.}
\label{FigPDOS}
\end{figure*}

 In order to understand the contribution of different orbitals to the electronic states, we have also performed the calculations of the total DOS and the partial DOS for puckered structures and the results are given in Fig.\ref{FigPDOS}. We find that the valence band structures of puckered structures have dominant contributions from the p orbitals  of  both group IV  and VI elements.  They additionally have very small contribution from the s orbitals.  The characteristic of p dominant valence states is commonly observed in monolayer honeycomb systems such as silicene, germanene, phosphorene and arsenene. Furthermore, it is also observed that  the valence states near the Fermi level have more contribution from the p orbitals of group VI atoms in comparison with that of group IV atoms. This is due to the fact that the number of valence electrons in group VI elements is higher than that of group IV elements.   However, we find that the carbon based systems such as CS, CSe, CTe show the opposite trend, where the states near the Fermi level has larger contribution from the p orbital of C. 
 
To understand the reason behind this deviation and also to check whether there is any charge transfer between the group IV and VI atoms, we have carried out the Bader charge analysis for all the puckered group IV-VI binary monolayers. The Bader charge is the amount of the charge in the Bader volume, which is defined so that the charge density takes a minimum on its surface. The Bader charge divides the total electronic charge of a system into its constituent atoms. 
 The results obtained from this analysis are  summarized in Table II.  It is clearly seen from Table II that there exists a net charge transfer between the constituent atoms. This confirms that there is an ionic contribution to the bonding between the atoms, which is in contrast to the purely covalent  bonding present between the atoms in the elemental monolayers.

\clearpage
 
\begin{figure}[!t]
\begin{center}
\includegraphics[width=0.45\textwidth]{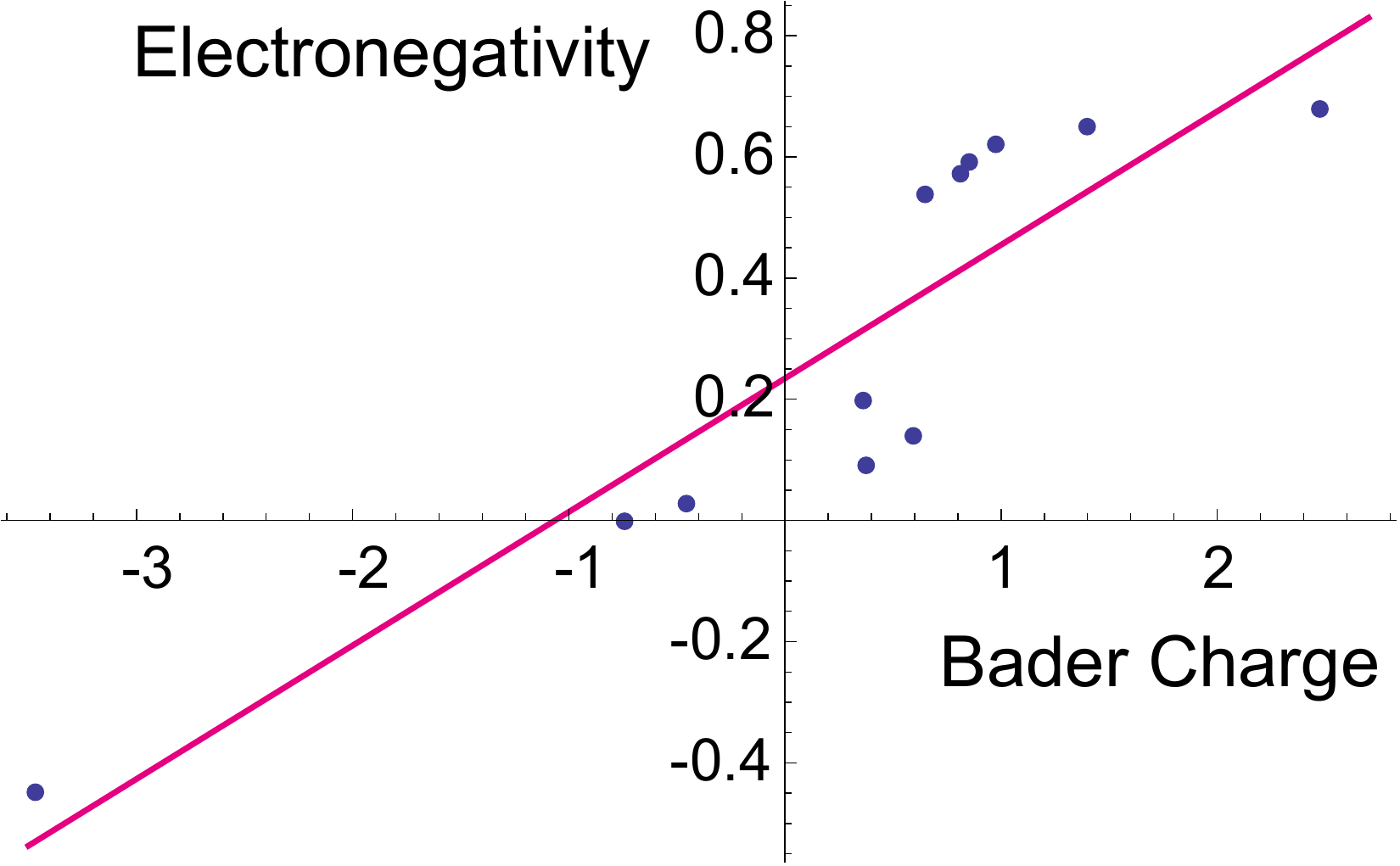}
\end{center}
\caption{(color online) The relationship between the Bader charge difference  and the electronegativity difference (in eV) between the $X$ and $Y$ atoms. They are almost proportional to each other.}
\label{FigBader}
\end{figure}

 It is interesting to note that the charge is getting transferred from X (group IV) to Y (group VI) atoms but with exception of CS, CSe and CTe.  The reason for the charge transfer from group IV  to  group VI atoms is due to the fact that the electronegativity for the group VI elements $\xi_X$ such as O, S, Se and Te (their respective values in Pauling scale are 3.44, 2.58, 2.55 and 2.1 eV) are higher that those of the group IV atoms $\xi_Y$ such as Si, Ge and Sn (their respective values 1.90, 2.01 and 1.96 eV).  Thus, the Si, Ge and Sn atoms happen to lose charges to O, S, Se and Te.  On the other hands, the electronegativity of carbon (2.55 eV)  is comparable to that of S, and Se  and higher than that of Te. Hence, there is some amount of charge getting transferred from S, Se and Te to C atoms. Thus,  based on the electronegativity of elements it is possible to explain the trend in the charge transfer between the constituent atoms in group IV-VI binary monolayers. We show the relationship between the Bader charge difference and the electronegativity difference in Fig.\ref{FigBader}. We find that they are to some extent proportional to each other.

Another important observation is that the band structures of group IV-VI binary monolayers possess many conduction band minima and valence band maxima. In several cases, these minima and maxima lie at the same momentum vectors. Thus, an electronic transition between these extrema can take place through optical absorption. We wish to note here that the energy differences between the fundamental indirect and direct band gaps for several group IV-VI binary monolayers in the puckered configuration are small. The values of indirect, direct band gaps and their difference are given in seventh, eighth and ninth columns of Table I, respectively.  Hence, there is a possibility of making these binary monolayers  to undergo a transition from indirect to direct band gap by suitable external influences.  In this case, we choose to apply mechanical strain, both compressive and tensile, on these structures and probe whether they can be converted into direct band gap materials since it is well known that the direct band gap semiconductors are preferred over the indirect band gap one for any optoelectronic device applications. Further, application of strain can be achieved by using a suitable substrate.

\begin{table}[!]
\begin{center}
\caption{Bader charge analysis for the group IV-VI binary monolayers: The charge inside the Bader volumes around X (group IV),  Y (group VI) atoms, the net charge transfer between them and the difference of the electronegativity ($\xi_Y-\xi_X$ ) of X and Y atoms.}
 \begin{tabular}{ccccc}
\hline
\hline
System	&	\multicolumn{2}{c}{Bader Charge on}		&	Charge Transfer	&	$\xi_Y-\xi_X$ \\ \cline{2-3}
XY	&	X	&	Y	&	From X to Y&	 (eV)	\\
\hline
\hline
CO	&	2.164	&	7.836	&	1.836	&	0.89	\\
SiO	&	1.643	&	8.357	&	2.357	&	1.54	\\
GeO	&	12.763	&	7.237	&	1.237	&	1.43	\\
SnO	&	12.709	&	7.291	&	1.291	&	1.48	\\
CS	&	4.454	&	5.547	&	-0.454	&	0.03	\\
SiS	&	1.519	&	8.481	&	2.481	&	0.68	\\
GeS	&	13.185	&	6.815	&	0.815	&	0.57	\\
SnS	&	13.020	&	6.980	&	0.980	&	0.62	\\
CSe	&	4.737	&	5.263	&	-0.737	&	0.00	\\
SiSe	&	2.608	&	7.392	&	1.392	&	0.65	\\
GeSe	&	13.351	&	6.649	&	0.649	&	0.54	\\
SnSe	&	13.146	&	6.855	&	0.855	&	0.59	\\
CTe	&	7.474	&	2.526	&	-3.474	&	-0.45	\\
SiTe	&	3.638	&	6.362	&	0.362	&	0.20	\\
GeTe	&	13.628	&	6.372	&	0.372	&	0.09	\\
SnTe	&	13.404	&	6.596	&	0.596	&	0.14	\\
\hline
\end{tabular}
\end{center}
\end{table}

\subsection{Influence of Mechanical Strain}

To probe the possibility of strain induced indirect-to-direct band gap transition in group IV-VI binary monolayers, we have carried out the band structure calculations of these systems, when they are under the influence of mechanical strain. We apply both compressive and tensile strains along the two lattice vectors ``a'' and ``b'' (See Fig.1 ).  The application of mechanical strain is simulated by constraining the lattice constant and relaxing the position of each atom during the geometric optimization. The amount of strain is represented as the change in the lattice constant from its fully optimized geometry  ($\pm \Delta a $ or $\pm \Delta b$) . The positive and negative signs indicate the tensile and compressive strains, respectively. We have performed the calculations for mechanical strain from 1 to 3 \%. 

\begin{figure*}[!t]
\begin{center}
\includegraphics[width=0.6\textwidth]{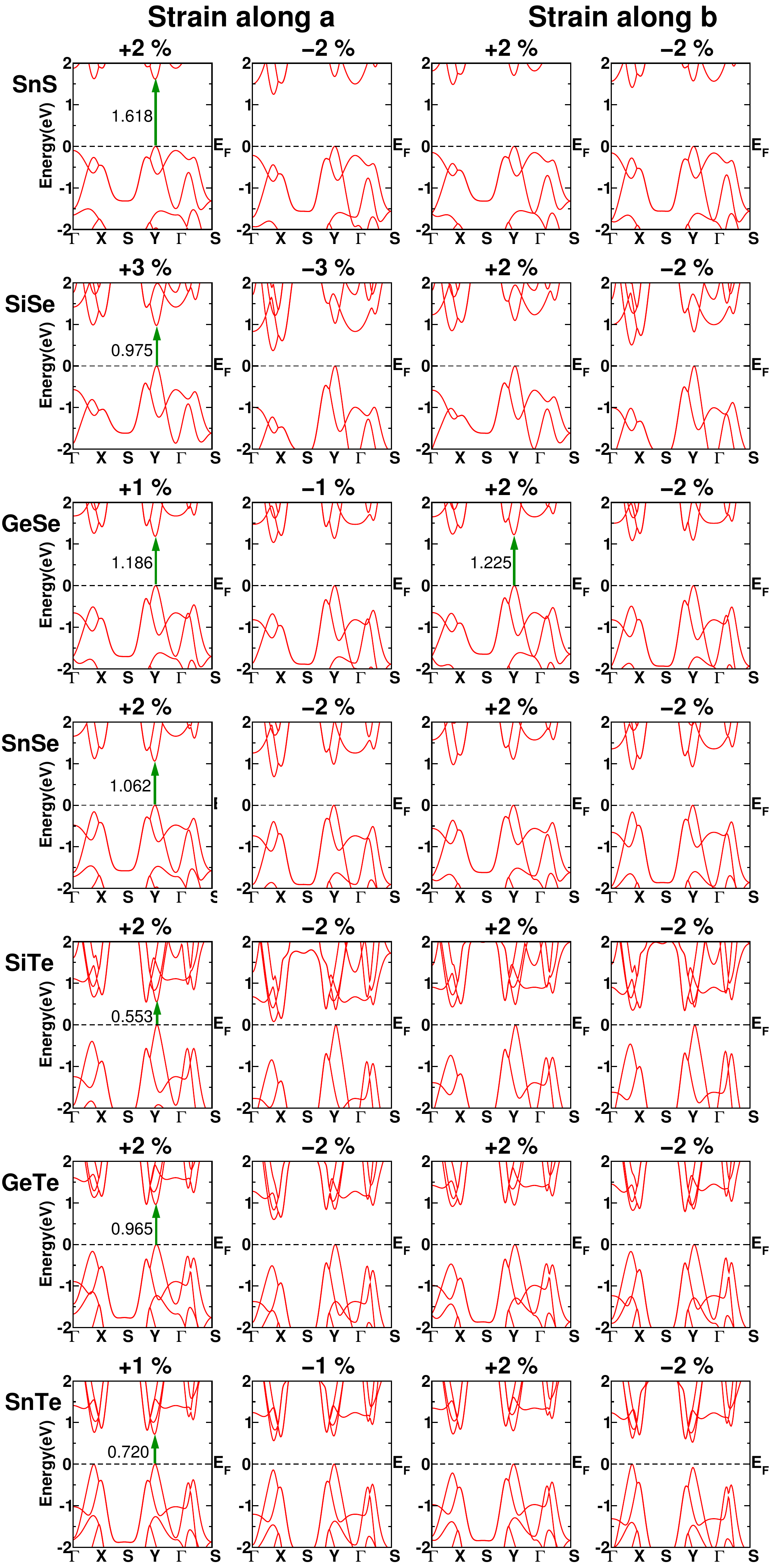}
\end{center}
\caption{(color online) Variation of the band structure of group IV-VI monolayers in the puckered configuration with small mechanical strains. Both compressive (negative) and tensile (positive) \textcolor{red}{strains} are applied along the lattice direction $a$ and $b$. The systems, mentioned above, undergo a indirect-to-direct band gap transition due to the application of very small strain ($\leq$ 3 $\%$). The values above sub-figure represent the amount of strain required to make this transition.  The vertical arrows (blue color) indicate the vertical transitions from the local maxima of the valence band to the local minima of the conduction band. Numerical values written near the arrows represent the corresponding band gaps.}
\label{FigBandStruc4}
\end{figure*}

Our calculations predict that the seven systems (SnS, SiSe, GeSe, SnSe, SiTe, GeTe and SnTe) undergo an indirect-to-direct gap transition by the application of  mechanical strain within this range. The results of  their band structures with strain along the ``a'' and ``b'' lattice directions are given in Fig.10.  The results indicate that the tensile strain along the ``a'' direction mainly induces an indirect-to-direct gap transition.  It is important to note that for many indirect band gap sulfides, selenides and tellurides, the conduction band edge emerges along the $\Gamma-X$ direction in the reciprocal lattice. The valence band edge emerges either at the $\Gamma$ point or along the $Y-\Gamma$ direction. The application of tensile (compressive) strain along the ``a'' lattice direction easily moves the conduction band (including the conduction band edge) along the $\Gamma-X$ direction away from (towards) the Fermi level, respectively.  However, the conduction band along another direction is nearly not affected by this strain. This leads to the situation where the conduction band minimum along the $Y-\Gamma$ direction (which is not the lowest point in absence of mechanical strain) becomes the conduction band edge. This causes the system to become a direct band gap semiconductor.

Different materials need different values of tensile strain. For GeSe and SnTe, a small tensile strain of 1\% is enough to make this transition, whereas 3\% of strain is required for SiSe. Tensile strain of 2\% is sufficient to make the remaining four systems (SnS, SnSe, SiTe and GeTe) undergo an indirect-to-direct gap transition. These differences of the magnitude of strain are naturally understood due to the fact that the energy difference between the indirect and direct gap is very small for GeSe ($0.015$) and SnTe ($0.033$) but large for SiSe ($0.287$).

On the other hand, it is observed that the application of strain along the ``b'' direction influences the conduction band in nearly the same manner in all the directions of the reciprocal lattice. The effect of  the tensile (compressive) strain along the ``b'' direction moves the conduction band towards (away from) the Fermi level in a nearly rigid fashion.  However, we have noted that GeSe becomes a direct band gap semiconductor due to the tensile strain along the ``b'' direction. Actually, in this case, the difference between the fundamental indirect and direct gaps is very small ($15$ meV). Thus,  small difference in the changes in the conduction band can cause the transition.  

Our calculations show that it is possible to make an indirect-to-direct band gap transition in as many as seven group IV-VI binary monolayers by the application of small mechanical strain of about $\leq$3 \%. These results suggest that group IV-VI binary monolayers are promising 2D materials, counterparts of phosphorene, for the  application in future optoelectronic devices.

\begin{table}[]
\footnotesize
\begin{center}
Experimentally available geometrical parameters and band gaps for the group IV-VI binary bulk materials (The superscripts show the references for the values listed here).
 \begin{tabular}{lccccc}
\hline
\hline

	&			\multicolumn{2}{c}{Lattice  Constant (\AA{})}&	\multicolumn{2}{c}{Band Gap (eV)}	\\  \cline{2-3}  \cline{4-5}
XY	&		a&	b	  &  	 Indirect	&	Direct 	\\

\hline
GeS	&		3.65\cite{Vaug}	&	4.30\cite{Vaug}		&1.58\cite{Vaug}	&	1.61\cite{Vaug}		&	\\
SnS	&		3.98\cite{Tan}	&	4.33\cite{Tan}		&	1.049\cite{Pare}	&	1.296\cite{Pare}		\\
GeSe	&	3.81\cite{Vaug}	&	4.37\cite{Vaug}	&	1.14\cite{Muk}	&	1.21\cite{Muk}	\\
SnSe	&		4.135\cite{Zhao}	&	4.44\cite{Zhao}		&	0.903\cite{Pare}	&	1.047\cite{Pare}		\\
\hline
\hline
\end{tabular}
\end{center}
\end{table}

\section{Conclusion}
We have studied in detail the geometric, energetic and electronic properties of group IV-VI binary monolayers (XY,  with X = C, Si, Ge, Sn;  Y = O, S, Se, Te) by employing DFT based calculations. For each material, we have considered three possible geometrical configurations, the puckered, buckled and planar structures.  Our calculations predict that among these three configurations, the puckered structure, as that of phosphorene, is the minimum energy configuration, whereas the planar structure is the least stable configuration.  Moreover, the  binding energy of the buckled configuration is very close to that of the puckered configuration. 

The electronic band structure calculations show that SiO and CSe in the puckered configuration are direct band gap semiconductors with gaps of 1.449 and 0.905 eV, respectively.  Interestingly,  CSe possesses the band structure quite similar to that of  phosphorene, which suggests that the electronic properties of the former will be similar to that of the latter. Hence, CSe can be considered as an alternate 2D material of phosphorene. The similarity between CSe and phosphorene can be understood by the fact that the electronegativity is identical for C and Se.  

All the remaining group IV-VI binary monolayers are found to be  indirect band gap  semiconductors. It is observed that these semiconducting monolayers have very small difference between their indirect and direct band gaps. Accordingly, by the application of very small mechanical strain ($\leq$ 3 $\%$),  it is possible to modify the semiconducting properties from indirect to direct band gap for as many as seven binary monolayers (SnS, SiSe, GeSe, SnSe, SiTe, GeTe and SnTe).  
Finally, we note that there are several experimental reports on Layered group IV-VI compounds such as SnS\cite{Wied,Tan,Pare,Rama}, SnSe\cite{Wied,Zhao,Eym,Pare}, GeS\cite{Wied,Vaug,LiACS}, GeSe\cite{Wied,Muk,Vaug,Xue}, GeTe\cite{Chat}. 
 Moreover, the results of geometrical parameters obained for group IV-VI monolayers match well with the experimentally obtained data for their corresponding bulk materials available in the literature. Thus, it is possible to obtain their monolayer counterparts by the exfoliation method.
In light of the existing literature combined with the present work, we expect that group IV-VI binary monolayers may well be promising 2D materials for future light-emitting diodes and solar cells.  

\section{Acknowledgment}
 C. K. and A. C.  thank Dr. P.A. Naik and Dr. P. D. Gupta for support and encouragement. C.K. and A.C. also thank the Scientific Computing Group, RRCAT for their support.  M. E. is very much grateful to Prof. N. Nagaosa for many helpful discussions on the subject.
M. E. thanks the support by JSPS KAKENHI Grants No. 25400317  and No. 15H05854.

\end{document}